\documentclass[useAMS,usenatbib,usegraphicx,twocolumn]{mn2e}

\usepackage{epsf}

\begin{document}
\newcommand{\msun}{M_{\odot}}
\newcommand{\kms}{\, {\rm km\, s}^{-1}}
\newcommand{\cm}{\, {\rm cm}}
\newcommand{\gm}{\, {\rm g}}
\newcommand{\erg}{\, {\rm erg}}
\newcommand{\kel}{\, {\rm K}}
\newcommand{\kpc}{\, {\rm kpc}}
\newcommand{\mpc}{\, {\rm Mpc}}
\newcommand{\seg}{\, {\rm s}}
\newcommand{\kev}{\, {\rm keV}}
\newcommand{\hz}{\, {\rm Hz}}
\newcommand{\etal}{et al.\ }
\newcommand{\yr}{\, {\rm yr}}
\newcommand{\gyr}{\, {\rm Gyr}}
\newcommand{\eq}{eq.\ }
\newcommand{\amunit}{\msun {\rm AU^2/yr}}
\def\arcsec{''\hskip-3pt .}
\def\deg{^{\circ}}

\def\gapprox{\;\rlap{\lower 3.0pt                       
             \hbox{$\sim$}}\raise 2.5pt\hbox{$>$}\;}
\def\lapprox{\;\rlap{\lower 3.1pt                       
             \hbox{$\sim$}}\raise 2.7pt\hbox{$<$}\;}

\newcommand{\figsizeFour}{5.5cm}

\newcommand{\figwidthSingle}{14.0cm}
\newcommand{\figwidthDouble}{7.50cm}

\newcommand{\figbig}{\figwidthSingle}
\newcommand{\figsmall}{\figwidthDouble}

\newcommand{\figappend}{\figwidthSingle}

\newcommand{\reqOne}[1]{Equation~(\ref{#1})}
\newcommand{\reqTwo}[2]{Equations~(\ref{#1}) and~(\ref{#2})}
\newcommand{\reqNP}[1]{Equation~\ref{#1}}
\newcommand{\reqTwoNP}[2]{Equations~\ref{#1} and~\ref{#2}}
\newcommand{\reqTo}[2]{Equation~(\ref{#1})-(\ref{#2})}
\newcommand{\rn}[1]{(\ref{#1})}
\newcommand{\ern}[1]{equation~(\ref{#1})}
\newcommand{\be}{\begin{equation}}
\newcommand{\ee}{\end{equation}}
\newcommand{\ff}[2]{{\textstyle \frac{#1}{#2}}}
\newcommand{\ben}{\begin{enumerate}}
\newcommand{\een}{\end{enumerate}}

\title[The stability radius from three body stability]
{Application of three body stability to globular clusters I: the stability radius}

\author[Gareth F. Kennedy]
  {Gareth F. Kennedy$^{1,2}$\thanks{Corresponding author email: gareth.f.kennedy@gmail.com}\\
      $^1$ National Astronomical Observatories of China, Chinese Academy of Sciences, Beijing 100012, China \\
      $^2$ Monash Centre for Astrophysics, Monash University, Clayton, Vic, Australia, 3800 
}

\maketitle

\begin{abstract}

The tidal radius is commonly determined analytically by equating the tidal field of the galaxy to the gravitational potential of the cluster. Stars crossing this radius can move from orbiting the cluster centre to independently orbiting the galaxy. In this paper the stability radius of a globular cluster is estimated using a novel approach from the theoretical standpoint of the general three-body problem. This is achieved by an analytical formula for the transition radius between stable and unstable orbits in a globular cluster.

A stability analysis, outlined by \citet{RoIoA}, is used here to predict the occurrence of unstable stellar orbits in the outermost region of a globular cluster in a distant orbit around a galaxy. It is found that the eccentricity of the cluster-galaxy orbit has a far more significant effect on the stability radius of globular clusters than previous theoretical results of the tidal radius have found. A simple analytical formula is given for determining the transition between stable and unstable orbits, which is analogous to the tidal radius for a globular cluster. The stability radius estimate is interior to tidal radius estimates and gives the innermost region from which stars can random walk to their eventual escape from the cluster. The timescale for this random walk process is also estimated using numerical three-body scattering experiments. 

\end{abstract}


\section{Introduction}

The tidal radius of a globular cluster (GC) is defined as the point at which stars will escape the cluster's potential well and become part of the galactic halo. It is typically calculated by considering the local equilibrium point in the potential between the cluster and the galaxy \citep[see][]{KingI,RWEGK2006}. The aim of this paper is to determine the boundary between tidally stable and unstable orbits in a cluster potential. An unstable orbit refers to a star orbiting inside the globular cluster that will eventually cross the tidal radius and escape the cluster. In terms of the removal of stars the stability boundary is analogous to the tidal radius of a globular cluster and is predicted using the stability analysis of \citet{RoIoA} as applied to GCs on eccentric galactic orbits and with arbitrary mass ratios.

For the purposes of applying the Mardling stability criterion, the star-cluster-galaxy system is approximated as follows. A star of mass $m_i$ orbits a particle representing the total cluster mass $M_C$ which itself orbits the galaxy, taken as a particle of mass $M_G$. Each of these particles is treated as a point mass so that the system can be considered as a general three-body problem. This means that we require a stability analysis that allows for: (1) mass ratios of $q_i = m_i/M_C \sim 10^{-5}$ and $q_o = M_G/M_C \sim 10^6$, (2) eccentric orbits, (3) large period ratios, (4) inclined orbits and (5) predicting instability on timescales of the order of ten GC-galaxy orbital periods. Each of these factors are found in this work to significantly alter the stability of a three-body system, so any stability criteria candidate must include all of these factors and be valid for the appropriate parameter ranges.

Existing stability criteria in the literature typically fail at least one of these requirements. For example \cite{EggletonKiseleva1995} empirically determine a stability criterion valid in the mass ratio range $1 < q_{i} < 100$ and $10^{-2} < q_{o} < 10^2$, which is not applicable here. More common is the assumption of coplanar and circular orbits \citep[e.g.][]{ECM2008} or low period ratios approached using Hill stability \citep[e.g.][]{BarnesGreenberg2007} or periodic orbits \citep{VH2005,Hadjidemetriou2006}. Fast Lypaunov indicators have been applied to general three-body planetary systems as a way of distinguishing between regular and chaotic orbits \citep{FLG1997,SSEPD2007}. However these require that the differential equations describing the system be numerically integrated for a few hundred times the outer period and then using the Fast Lypaunov exponents to determine the chaotic likelihood of a given orbit. The focus of this paper is to predict the chaotic regions without requiring numerical experiments, only using these to validate the stability analysis.

Our approach is to use the mean motion resonance overlap to predict the occurrence of unstable orbits. \cite{Wisdom1980} gives a good introduction to the application of resonance overlap criterion to the restricted three-body problem. The interested reader is also referred to the review by \cite{Chirikov1979}. \cite{Wisdom1982} and more recently \cite{QF2006} give examples of how resonance overlap theory successfully predicts unstable regions that are tested with numerical results; although these studies were limited in scope by low period ratio and coplanarity respectively. 
\cite{MudrykWu2006} also use resonance overlap to predict the ejection of a single planet from a binary star system, however they too are limited to coplanar systems and use different mass ratios to those required here. 
The overlap of nonlinear secular resonances was recently examined by \cite{WuLithwick2011} and \cite{LithwickWu2011}, but this too is not applicable here since we are interested in timescales of a few GC-cluster orbits, so mean motion resonances are important while secular resonances do not have time to occur. The only known stability analysis that meets all of the previous requirements is the Mardling stability criterion presented in \citet{RoIoA}, with additional terms for inclined orbits. This is based on the theory of resonance overlap producing chaotic orbits and is discussed in depth in Section~\ref{MSC}.

A recent comparison between stability criteria in the context of stellar mass triples was conducted by \cite{ZKO2010}; their tests found that the best stability criteria were from \cite{ValtonenEtAl2008} and \cite{Aarseth2003}. The stability criterion used in \cite{Aarseth2003} does not include any inclination dependence and is not accurate for the high mass ratios in the galaxy-GC system. Therefore in Section~\ref{stability_test} we compare the Mardling stability predictions and the stability criterion of \cite{ValtonenEtAl2008} to numerical experiments of inclined orbits.

By approximating the cluster potential as a point mass in the stability analysis we are assuming that escaping stars spend most of their time in the outermost regions of the cluster prior to escape. A simplified treatment of the galactic potential using a point mass of $10^{11} \msun$ is adopted for analytical convenience. The choice of $10^{11} \msun$ is based on a circular velocity of 220 km/s at the galactocentric distance of the sun. This assumption breaks down at large distances where the halo potential is closer approximated by an isothermal sphere \citep[e.g.][]{IrrgangEtAl2013}. The effect of an isothermal sphere and more realistic galactic potentials are examined in an upcoming paper.

Simplifying the star-cluster-galaxy system to a single three-body problem ignores the effect of mutual interactions between stars in the cluster, overlooking two-body relaxation. This means that in a real cluster, stars will be able to diffuse over the predicted tidal radius and escape the cluster from orbits that were initially in stable regions. Generally the outer regions of the cluster have very long timescales compared to the timescale of a star's escape from the cluster. A more detailed timescale comparison, including where the assumption of two-body relaxation being negligible is valid, is undertaken in Section~\ref{gc:valid}.

This paper is structured in the following way. A brief overview of the Mardling stability criterion is presented in Section~\ref{MSC}. Validation of this stability analysis for inclined orbits using numerical orbital integrations is presented in Section~\ref{stability_test} and investigation of the associated escape timescales in made in Section~\ref{gc:valid}. The stellar orbits occurring inside a GC, particularly the distribution function for orbital eccentricity, is calculated in Section~\ref{GCStableRegions}. Section~\ref{GCRTidal} applies the stability boundary to calculate the stability radius for a case study globular cluster of mass $M_c = 10^{6} \msun$ and this is compared to previous theoretical work from the literature in Section~\ref{Comparison}. Conclusions and a discussion of the observational and numerical simulation implications of this work are summarised in Section~\ref{Conc}.

\section{Mardling stability criterion}
\label{MSC}

We first summarise the Mardling stability criterion (MSC) which is based on resonance overlap and outline the algorithm used to determine the stability of any three-body system. The criterion for the MSC is given in \citet{RoIoA}, and interested readers are referred to this study for details which have been omitted here. The criterion below covers the more general case where the inner and outer orbits are relatively inclined by an angle $I$ (Mardling, in preparation). The inner orbit refers to the orbit of the binary composed of point masses $m_1$ and $m_2$, while the outer orbit refers to the orbit of $m_3$ with the centre of mass associated with the combined mass $m_{12} = m_1 + m_2$. 

Predicting unstable systems via resonance overlap has a long history, beginning with \cite{Chirikov1959} who found that a motion in a system will change from deterministic trajectories to chaotic motion once two resonances overlap. This occurs when $K \sim (\Delta \sigma_1 + \Delta \sigma_2)/\Omega_{12} > 1$ where $\Delta \sigma_i$ refers to the unperturbed width of resonance $i$ and $\Omega_{12}$ is the distance between resonances. In this work $\Delta \sigma_n$ is the width of the $n:1$ mean motion resonance and $\Omega_{12} = 1$ since it is the distance in period ratio space between the $n:1$ and $n+1:1$ resonances.

We will use the common terminology of a system being in resonance if the ratio of the outer to inner periods ($\sigma = T_o/T_i$) is within a distance $\Delta \sigma$ of a particular ($n:n'$) resonance. 
When a system has initial conditions such that it is librating in one resonance and a neighbouring resonance is sufficiently close then it can force it enough to make it circulate and jump into another resonance (see also \citeauthor{Wisdom1980} \citeyear{Wisdom1980} and \citeauthor{Laskar1990} \citeyear{Laskar1990} for an application to secular resonances). This jumping between resonances can occur at any time and means that two three-body systems with initially very close conditions will quickly diverge once one jumps to a different resonance, leaving the other behind. This sensitivity to changes in the initial conditions is characteristic of chaotic motion and is used by the MSC to predict unstable systems.

For a range of orbits, as exists for stars in globular clusters, the inclination between the inner and outer binary orbits is not restricted to the coplanar case. The method given in \citet{RoIoA} for calculating the resonance widths does not include the effect of the relative inclination $I$. The effect of inclination on the resonance width is under development (Mardling, in preparation) and we present here a summary of the inclination factors relevant to this paper. In addition to the inclination, the phase of the orbit also changes over time due to the non-Keplerian motion in the cluster centre (see \reqNP{PlummerPotentialUS} for the GC potential model). For simplicity the maximum possible resonance width is adopted which is equivalent to taking the resonance angle as zero.

From the formulation given in \citet{RoIoA} and including the inclination terms from Mardling (in preparation) the resonance width is given by
\begin{equation}
\label{th_ResWidth}
\Delta \sigma_{mnn'} = 2 \sqrt{\mathcal{A}_{mnn'}}
\end{equation}
where for the leading quadrupole term ($l=2$) in the spherical expansion of the disturbing function one can write
\begin{equation}
\mathcal{A}_{mnn'} = \max(\mathcal{A}_{2n12}, \mathcal{A}_{2n10}, \mathcal{A}_{2n1-2}, \mathcal{A}_{0n10})
\label{th_CurclyAInc}
\end{equation}
since $m = -l, 0, l$ and limiting to resonances where $n' = 1$ and $n$ is a positive integer\footnote{\cite{MudrykWu2006} also found that mean motion resonance overlaps for high order resonances (e.g. 50:3) can be neglected for high period ratios.} then
\begin{eqnarray}
\mathcal{A}_{mn n'm'} & = & -6 c_{2m'}^2 s_{n}^{(2m')}(e_i) F_n^{(2m)}(e_o) \gamma_{2mm'}(I) \nonumber \\
& \times & \left[ \frac{m_3}{m_{123}} + n^{2/3} \left( \frac{m_{12}}{m_{123}} \right)^{2/3} \left( \frac{m_1 m_2}{m_{12}^2} \right) \right],
\label{th_AInc}
\end{eqnarray}
with $c_{22}^2 = 3/8$ and $c_{20}^2 = 1/4$. The approximate dependence on the inner eccentricity is given by the functions
\begin{eqnarray}
s_{1}^{(20)}(e_i) & \approx & \frac{e_i}{9216}\left( -9216 + 1152 e_i^2 - 48 e_i^4 + e_i^6 \right) \nonumber \\
s_{-1}^{(22)}(e_i) & \approx & -\frac{e_i^3}{15360}\left( 4480 + 1880 e_i^2 + 1091 e_i^4 \right) \nonumber \\
s_1^{22}(e_i) & \approx & -3e_i + \frac{13}{8}e_i^3 + \frac{5}{192} e_i^5 - \frac{227}{3072}e_i^7
\label{th_ResFn_ei}
\end{eqnarray}
which is valid for $0 \leq e_i \leq 1$. The errors between this approximate expression and the exact integral expression are less than 1\% for $e_i < 0.8$ and $<$ 0.1\% for $e_i < 0.63$ \citep{RoIoA}. The dependence of \reqOne{th_AInc} on the outer eccentricity is approximated by the asymptotic expression
\begin{eqnarray}
F_n^{(22)}(e_o) & \approx & \frac{8\pi \sqrt{2\pi}}{3} \frac{(1-e_o^2)^{3/4}}{e_o^2} n^{3/2}e^{-n \xi(e_o)} \nonumber \\
F_n^{(20)}(e_o)  & \approx & \frac{1}{\sqrt{2 \pi n}} \left( 1-e_o^2 \right)^{-3/4} e^{-n \xi(e_o)}
\label{th_ResFn_eo}
\end{eqnarray}
where
\begin{equation}
\xi(e_o) = \left( \mathrm{Cosh}^{-1}(\frac{1}{e_o}) - \sqrt{1-e_o^2} \right). 
\label{th_xi_eo}
\end{equation}
The relevant inclination factors are 
\begin{eqnarray}
\label{th_incStart}
\gamma_{222}(I) & = & \frac{1}{4} \left( 1 + \cos(I) \right)^2 \\
\gamma_{220}(I) & = & \sqrt{\frac{3}{8}} \sin^2(I) \\
\gamma_{22-2}(I) & = & \frac{1}{4} \left( 1 - \cos(I) \right)^2 \\
\gamma_{200}(I) & = & \frac{1}{2} \left( 3 \cos^2(I) - 1 \right).
\label{th_incEnd}
\end{eqnarray}
The MSC, including these inclination terms, has been successfully used in n-body codes to avoid integrating stable triple systems for over a decade \citep[e.g.][]{Aarseth2003}.

As a system evolves on a secular timescale the resonance widths vary and are at their maximum when $e_i$ is maximum. Therefore the inner eccentricity in \reqOne{th_ResFn_ei} needs to be modified for induced eccentricity in the inner orbit due to the outer orbit and the secular octopole term. The maximum eccentricity that can be dynamically induced in the inner eccentricity by the outer orbit following one passage is given by
\begin{equation}
\label{th_einduced}
e_i^{ind}=\left[ e_i(0)^2 - 2 \beta_n e_i(0) \sin(\phi_{2n1}) + \beta_n^2 \right]^{1/2}
\end{equation}
where $e_i(0)$ denotes the initial inner eccentricity, and
\begin{equation}
\beta_n = \frac{9 \pi}{2 n} \left( \frac{m_3}{m_{123}}\right) \left(1 - e_o \right)^3 F_n^{(22)}(e_o)
\end{equation}
where $F_n^{(22)}(e_o)$ is given by \reqOne{th_ResFn_eo}. As this occurs over a single outer orbital period then the associated timescale is $\tau_{ind} \sim T_o$.
The eccentricity correction due to the secular octopole term associated with the $n=n'=0$ resonance, which is non-zero if $m_1 \neq m_2$, is
\begin{equation}
\label{th_eoctopole}
e_i^{oct} = \left\{ \begin{array}{cl}
(1+\alpha) e_i^{eq}, & \alpha \leq 1 \\
e_i(0)+2 e_i^{eq},   & \alpha > 1 
\end{array}
\right.
\end{equation}
where 
\begin{equation}
\alpha = \left| 1 - \frac{e_i(0)}{e_i^{eq}} \right|
\end{equation}
and in the limit $e_i << 1$
\begin{equation}
e_i^{eq} = \frac{ (5/4) e_o m_3 (m_1 - m_2) (a_i/a_o)^2 \sigma (1-e_o^2)^{-1/2} } { \left| m_1 m_2 - m_{12} m_3 \sqrt{a_i/a_o} \sigma \sqrt{1-e_o^2} \right| }.
\end{equation}
The characteristic timescale for the secular resonance can be determined by solving the time derivative for the inner eccentricity given in equation~48 of \cite{Mardling2013} and taking the timescale as the oscilation period $e_i$. This timescale is then
\begin{equation}
\tau_{oct} = T_o \frac{16}{15 \sqrt{7} } \left( \frac{M_G}{M_C} \right)^{1/3} \sigma^{5/3} \frac{\left( 1 - e_o^2 \right)^{5/2}}{e_o}
\end{equation}
which is $\sim 10^3 T_o$ for typical values used later and is consistent with the expectation of $\gapprox 10^3 T_i$ from \cite{MD1999}.

For inclined systems an additional secular effect is important, this is known as the Kozai effect \citep[see][]{InnanenEtAl1997} and involves a relationship between the eccentricity and inclination such that the maximum eccentricity induced by the Kozai mechanism is
\begin{equation}
\label{th_eKozai}
e_K = \sqrt{ \frac{1}{6} \left| Z + 1 - 4 A^2 + \sqrt{ D } \right| }
\end{equation}
where
\begin{eqnarray}
A & = & \cos I \sqrt{1 - e_i(0)^2} \\
Z & = & (1 - e_i(0)^2)(1+\sin^2I) \nonumber \\
 & + & 5 e_i(0)^2 (\sin \varpi \sin I)^2
\end{eqnarray}
and
\begin{equation}
D  = 16 A^4 - 20 A^2 - 8 A^2 Z - 10 Z + Z^2 + 25,
\end{equation}
which all depend on the initial eccentricity of the inner binary, $e_i(0)$, the initial relative inclination between the inner and outer orbits, $I$, and $\varpi = \varpi_o - \varpi_i$. The timescale for the Kozai cycle from \cite{InnanenEtAl1997} and put in a more convienient form is 
\begin{equation}
\tau_K = \frac{M_C}{M_G} T_o \sigma \left( 1-e_o^2 \right)^{3/2}
\end{equation}
which gives $\tau_K \sim 10^{-4} T_o$ for typical values used later.

The eccentricity determined by \reqOne{th_eKozai} is the maximum possible eccentricity that comes out of this Kozai cycle that gives the maximum resonance width (Mardling, in preparation). This means that the maximum eccentricity induced by the Kozai mechanism ($e_K$) must also be included in the inner eccentricity functions given in \reqOne{th_ResFn_ei}. This is achieved by replacing $e_i$ in these equations with the theoretical maximum inner eccentricity given by
\begin{equation}
\label{th_eincorrect}
e_i = \max ( e_i^{ind}, e_i^{oct}, e_K )
\end{equation}
where $e_i^{ind}$ is the induced eccentricity due to the outer binary orbit (\reqNP{th_einduced}) and $e_i^{oct}$ is the eccentricity correction due to the octopole term (\reqNP{th_eoctopole}). As the Kozai mechanism relates the eccentricity and inclination then the inclination used in \reqTo{th_incStart}{th_incEnd} must be replaced by the maximum possible inclination over a Kozai cycle ($I_K$). The maximum inclination is given by
\begin{eqnarray}
\label{inc1}
\cos(I_K) & = & \frac{A}{\sqrt{1-e_K}} \\
\sin(I_K) & = & \sqrt{1-\cos^2I_K}.
\label{inc2}
\end{eqnarray}
Each resonance width is calculated using the maximum $e_i$ (\reqNP{th_eincorrect}) and inclination $I_K$ (\reqTwoNP{inc1}{inc2}) together with \reqOne{th_eoctopole}. For the application to GC orbits in a galactic tidal field a given set of system parameters ($m_1$, $m_2$, $m_3$, $e_o$ and $R_p$) are fixed and the stability for each particular set of $e_i(0)$, $\sigma$ and inclination $I$ is determined.

From timescale arguments only the induced eccentricity ($\tau_{ind} \sim T_o$) and the Kozai eccentricity ($\tau_K << T_o$) will be significant when applied to GCs. Secular evolution can safely be ignored since the timescale is $\tau_{oct} \sim 10^3 T_o$, in addition to this the magnitude of $e_{oct}$ is far smaller than the other eccentricity effects for all parameters of interest here. The stability criterion updated with the new inclination terms for the masses relevant to this study is tested in the next section.

\section{Testing stability of inclined systems}

\label{stability_test}

In this section numerical orbital integrations are used to test the stability and timescale of escape of the stellar particle from the cluster potential for a range of period ratio $\sigma$, stellar orbital eccentricity $e_i$ and relative inclinations $I$. In all cases the system is set up using the previously stated convention, i.e. the inner binary is composed of a stellar particle of mass 1 $\msun$ orbiting a cluster particle of mass $10^6 \msun$ which itself orbits a galaxy particle of mass $10^{11} \msun$. For direct comparison with the stability analysis all masses are assigned point mass potentials and orbital elements are calculated assuming Keplerian motion.

\begin{figure}
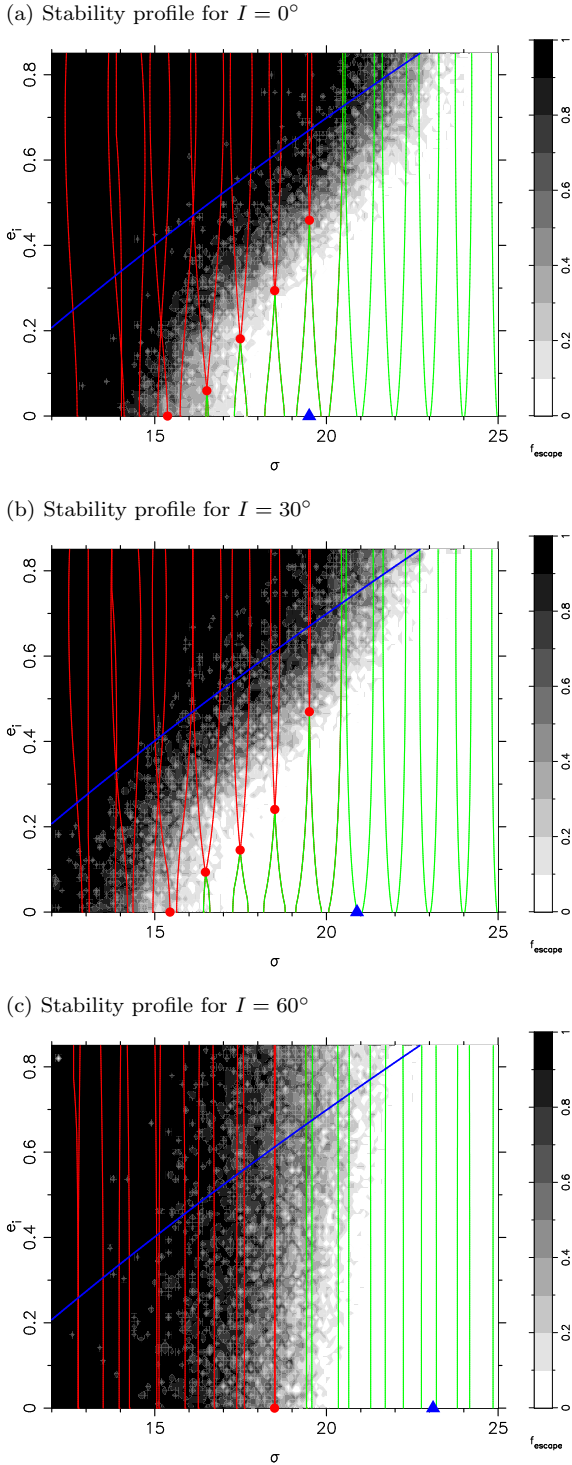

\begin{centering}$\begin{array}{c}
\multicolumn{1}{l}{\mbox{(a) Stability profile for $I=0\deg$}}\\[-0.1cm]
\includegraphics[height=\figsmall,angle=270]{ResOverlap_FN_I000.ps}\\
\multicolumn{1}{c}{\mbox{}}\\
\multicolumn{1}{l}{\mbox{(b) Stability profile for $I=30\deg$}}\\[-0.1cm]
\includegraphics[height=\figsmall,angle=270]{ResOverlap_FN_I030.ps}\\
\multicolumn{1}{c}{\mbox{}}\\
\multicolumn{1}{l}{\mbox{(c) Stability profile for $I=60\deg$}}\\[-0.1cm]
\includegraphics[height=\figsmall,angle=270]{ResOverlap_FN_I060.ps}
\end{array}$
\par\end{centering}
\caption[Stability of inclined orbits]
{Fraction of stars escaping beyond $2 R_K$ from the cluster centre as determined from numerical results of a GC-galaxy orbit with $e_o=0.5$ and mass ratios of $m_2/m_1 = 10^{-6}$ and $m_3/m_1 = 10^5$. Resonance boundaries, shown in red when overlapping and green otherwise, are calculated using the Mardling formulation given by \reqTo{th_ResWidth}{inc2} and are significantly more successful at predicting the escape of stars than the King radius (blue line, \reqNP{rtidal_king1}) and the Valtonen stability criterion (blue triangle, \reqNP{ValtonenCrit}).}
\label{gc_SB_Inclined}
\end{figure}

The numerical three-body orbits were solved numerically using a Bulirsch-Stoer integrator \citep{NR86} for 1 Gyr or until one of the particles escaped the system. The computations were carried out for $e_o = 0.5$ and with $I=0\deg$, $30\deg$ or $60\deg$ for a grid of $0 \le e_i \le 0.85$ with $\Delta e_i = 0.01$ and $12 \le \sigma \le 25$ with $\Delta \sigma = 0.1$ making a total of 11266 simulations per inclination\footnote{Simulations with $e_i > 0.85$ required extremely short time-steps and therefore very long computational times. While simulations run with poor time resolution gave spurious results in favour of unstable orbits. For these reasons simulations with $e_i > 0.85$ are not included here, note that this choice does not affect the determination of the stability boundary nor any of the results presented hereafter.}. 
In addition each of these simulations consisted of 10 random realisations of the relative initial phase of the orbit to get a fraction of unstable orbits, this is equivalent to 860 realisations per $\sigma$ value after averaging over $e_i$ (see below). Simulations are run until the star escapes beyond two times the King radius, which is given by \citep{KingI}
\begin{equation}
R_K = 0.7 R_p \left( \frac{M_C}{M_G} \right)^{1/3} \left( 3 + e_o \right)^{-1/3}
\label{rtidal_king1}
\end{equation}
and is based on the Jacobi radius, with a correction factor of 0.7 fitted to numerical results. The resulting fraction of orbits where one body escapes to $2 R_K$ as a function of the period ratio ($\sigma$) and the inner eccentricity ($e_i$) is shown in Figure~\ref{gc_SB_Inclined}. Note that one gets nearly identical results with a numerical test for chaos such as the maximum separation in semi-major axis of initially nearby orbits, similar to the test by \cite{Wisdom1980}. We have chosen to focus on the fraction of stars that escape the cluster since the timescale for the star to escape beyond the tidal radius is also of interest (see below). All simulations were conducted over many months using the computational facilities of the Monash eScience and Grid Engineering Laboratory ({MeSsAGE} Lab) \citep{NIMRODG}.

Predictions of two stability criteria were tested against the numerical results for the fraction of escaping stars shown in Figure~\ref{gc_SB_Inclined}. These being the MSC discussed in the previous section and the stability criterion from \cite{ValtonenEtAl2008} which \cite{ZKO2010} found to be the most accurate stability criterion of those they tested. Orbital inclination effects are included in the \cite{ValtonenEtAl2008} stability criterion, however there is no treatment of the eccentricity of the inner orbit ($e_i$). Rewriting their stability criterion for the period ratio gives
\begin{equation}
\sigma_V = 3^{3/2} \left(1-e_o \right)^{-21/12} \left( \frac{7}{4} + \frac{1}{2} \cos I - \cos^2 I \right)^{1/2}
\label{ValtonenCrit}
\end{equation}
where all quantities have been defined previously. It is worth noting that we are using their criterion outside of the mass range that it was originally tested in, and so do not expect it to be as accurate as the MSC. The critical period ratio given by $\sigma_V$ is indicated by a blue triangle on the $\sigma$-axis for each inclination in Figure~\ref{gc_SB_Inclined}. The King radius (\reqNP{rtidal_king1}) is also shown as a blue curve in these figures where all orbits above the curve are expected to escape. The King radius is not a good predictor of escaping stars for high inclination orbits as it is based on a simple coplanar analysis, this discrepancy is seen for high $\sigma$ with $I=60\deg$.

The resonance widths for each period ratio, as calculated by MSC using \reqOne{th_ResWidth}, are shown in Figure~\ref{gc_SB_Inclined} as red curves when they overlap and green curves otherwise. The points where two resonances overlap (red points) mark out the predicted boundary between unstable (low $\sigma$) and stable (high $\sigma$) orbits. Unstable orbits are expected to produce stars that escape the cluster. As expected, the MSC gives a much more accurate and detailed prediction of the occurrence of escaping stars compared to the stability criterion of \cite{ValtonenEtAl2008} for all the tested inclination values of $I=0\deg$, $30\deg$ and $60\deg$. 
Note that the steepness of the stability transition region increasingly becomes more `vertical' as the inclination increases, which was also seen in the context of a binary star system encountering a third body on an inclined parabolic orbit by \cite{Donnison2006}. 

The purpose of this section was to establish that the Mardling stability criterion is valid for inclined orbits. The MSC was found to be an excellent predictor of unstable regions for the inclined orbits over a wide range of period ratio, inner eccentricity and phase. The timescale for a star to escape the cluster is further examined in Section~\ref{gc:valid} in the context of which GC orbits one would expect three-body instability to be a significant factor.

\section{Stellar orbits in globular clusters}

\label{GCStableRegions}

The MSC consists of a stability analysis to determine particular orbital configurations of the three masses which are unstable to the escape of one of the masses. In the case of a star-cluster-galaxy system the only energetically possible end state for such an unstable configuration is the escape of the star from the cluster. Therefore stars on orbits that are found to be unstable are predicted to eventually escape the cluster. Before applying this to stars in a GC the distribution function of orbital elements, particularly the eccentricity, must be known.

Once again point mass potentials are used for the galaxy and cluster, which is valid for the star-cluster and cluster-galaxy distances of interest here. The three body system is then described by a star $m_i = 1$ M$_{\odot}$, the globular cluster $M_C = 10^{6} \msun$ and the galaxy $M_G = 10^{11} \msun$. In the notation of the stability analysis in Section~\ref{MSC} the orbit of the star-cluster is referred to as the inner orbit composed of $m_1 = M_C$ and $m_2 = m_i$, and the orbit of the cluster-galaxy as the outer orbit (where $m_3 = M_G$). 

\begin{figure}
\begin{centering}\resizebox{\figsmall}{!}{\includegraphics[angle=0]{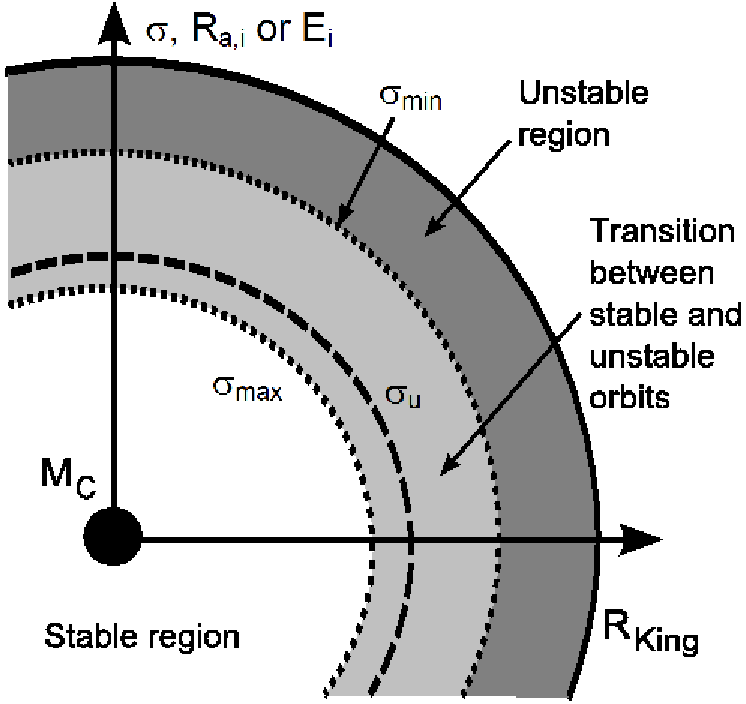}} 
\par\end{centering}
\caption[Conceptualisation of the stability of stellar orbits in a globular cluster]
{Conceptualisation of the stability of stellar orbits in a globular cluster. The distance from the cluster centre associated with the transition from unstable (dark shading) to stable orbits (unshaded inner region) is indicated by the ratio of outer to inner periods $\sigma_u$. The region where orbits can be found in either unstable or stable configurations is shown as a light shading between $\sigma_{min}$ and $\sigma_{max}$. The cluster is truncated at the maximum theoretical tidal radius which is equivalent to $R_{King}$ given by \reqOne{rtidal_king1}.}
\label{gc_3body_concept}
\end{figure}

Using the MSC a boundary between predominately unstable and stable orbits is sought, separated by a period ratio $\sigma_u$. The transition between the unstable exterior of the cluster and stable interior is characterised by two additional period ratio values $\sigma_{min}$ and $\sigma_{max}$. The difference between $\sigma_{max}$ and $\sigma_{min}$ represents the width of the transition region and can be used to estimate a maximum and minimum tidal radii range for globular clusters. A conceptualisation of the different stability regions for stellar orbits within a cluster is shown in Figure~\ref{gc_3body_concept}.

To determine the stability boundary inside a cluster we need to average over the eccentricity and inclination for the star-cluster orbit. In the case of the relative inclination between the star-cluster orbit and the cluster-galaxy orbit this is trivial, assuming a $\cos I$ distribution. In addition the period ratio, which requires an orbital period for the star-cluster, can be simply determined by assuming a point mass potential. This is valid in the outer regions of the cluster where the stability boundary occurs. However the eccentricity distribution is not so straight forward.

\label{EccDist}

To estimate a realistic eccentricity distribution for stars inside a globular cluster a simple simulation of $N=10^4$ particles in a non-Keplerian potential is computed. This number of particles was found to give sufficient resolution for the required distribution, while being computationally efficient.

A globular cluster is modelled using a Plummer sphere with gravitational potential given by \citep{BinneyTremaine1987}
\begin{equation}
\Phi=\frac{-GM_{C}}{\sqrt{b^{2}+r^{2}}}
\label{PlummerPotentialUS}
\end{equation}
where $G$ is the gravitational constant, $M_{C}$ is the mass of the cluster, $r$ is the radial distance, and $b$ is a softening parameter chosen to describe the compactness of the cluster. For $b = 0$ the Plummer potential reduces to the potential for a point mass of mass $M_C$. For simplicity a value of $b=1$ is assumed from herein. The physical scale for $b=1$ is approximately $0.77$ times the half-mass radius so that the potential is close to Keplerian at a few $R_{1/2}$, which is well inside the tidal radius for typical GC-galaxy orbits in the Milky Way.

Numerically modelling such a cluster requires the $N$ particles to be distributed such that the combined gravitational potential is equivalent to the Plummer potential, and the velocities are distributed by the equations given in \citet{AHW1974}. The initial distribution of particles is achieved using the mass enclosed within a particular radius along with von Neumann's rejection technique for random sampling from a distribution to determine the velocities. Here the same procedure as that of \citet{AHW1974} is followed, except for the alterations made to account for the cluster compactness parameter $b$ and for truncating the cluster at the King radius (\reqNP{rtidal_King}). Further technical details of this method can be found in \citet{MyThesis}.

\begin{figure}
\begin{centering}\resizebox{\figsmall}{!}{\includegraphics[angle=270]{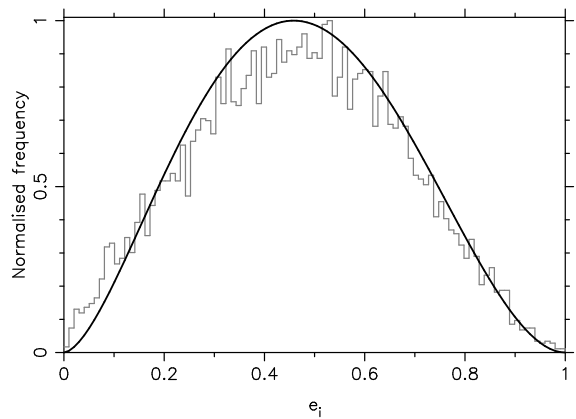}} \par\end{centering}
\caption
{The distribution of eccentricities for stellar orbits in a Plummer sphere with $b = 1$ and using an isotropic velocity distribution. A Beta distribution is fit to the data of the form given in \reqOne{gc_eccdistfn} and is shown as a grey curve for the best fit. The eccentricity distribution has a mean of $\bar{e_i} = 0.47$ for both the data and the fitting function.}
\label{gc_distEcc} 
\end{figure}

To calculate the orbit of each particle a Bulirsch-Stoer integrator \citep{NR86} was used until a minimum and maximum distance could reliably be determined, which was then used to give an eccentricity. The distribution of eccentricities for particle orbits in a cluster is shown in Figure~\ref{gc_distEcc}. Note that this eccentricity distribution is consistent, in terms of the shape and average value, with a full N-body integration conducted for a globular cluster using $\phi$-GRAPE \citep{HarfstEtAl2007}.

An integrable function that fits the eccentricity distribution in Figure~\ref{gc_distEcc} was sought, and the Beta distribution was found to provide the required fit. The Beta distribution is given by
\begin{equation}
f(e_i) = \frac{1}{B(\alpha,\beta)} e_i^{\alpha-1} \left( 1 - e_i \right)^{\beta - 1}
\label{gc_eccdistfn}
\end{equation}
where $B(\alpha,\beta)$ is the Beta function and the mean of the distribution is given by $\alpha/(\alpha+\beta)$. For the fitting function plotted as a grey curve in Figure~\ref{gc_distEcc}, $\alpha = 2.691732$ and we fix $\beta = 3$ for mathematical convenience. This value of $\alpha$ ensures the same mean eccentricity value of $\bar{e_i} = 0.47$ for the fitting function as for the distribution. The cumulative probability distribution for \reqOne{gc_eccdistfn} with $\beta = 3$ is
\begin{equation}
F(e_i) = \frac{1}{B(\alpha,\beta)} \left( \frac{1}{\alpha} e_i^\alpha - \frac{2}{\alpha+1}  e_i^{\alpha+1} + \frac{1}{\alpha+2}  e_i^{\alpha+2} \right)
\label{gc_CummEcc}
\end{equation}
where we can take advantage of the relationship between the Beta and Gamma functions to write
\begin{equation}
B(\alpha,\beta=3) = \frac{2}{\alpha \left( \alpha + 1 \right) \left( \alpha + 2 \right) }
\end{equation}
which is used when averaging over the eccentricity distribution to derive the fraction of unstable orbits in the next section.

Before moving onto the application of the stability boundary to a simple globular cluster model it is worth noting that the final analysis from Sections~\ref{GCRTidal} and~\ref{Comparison} was repeated with $2 < \alpha < 10$ and $\beta=3$. This is equivalent to changing the $f(e_i)$ distribution to become more dominated by radial orbits. It is important to test this since the effective value of $\alpha$ increases with decreasing $\sigma$ (i.e. increasing distance from the cluster centre), since distantly orbiting stars must have increased eccentricities to remain energetically bound to the cluster. It was found that the stability boundary does shift but the maximum relative error in any $\sigma$ value was $10\%$ with errors typically less than $5\%$, which translates into a relative error of less than $5\%$ for all radii. This means that the tidal radius predictions made in this paper are not strongly affected by the adopted eccentricity distribution.

\section{Application to globular clusters and approximating the stability radius}
\label{GCRTidal}

\begin{figure}
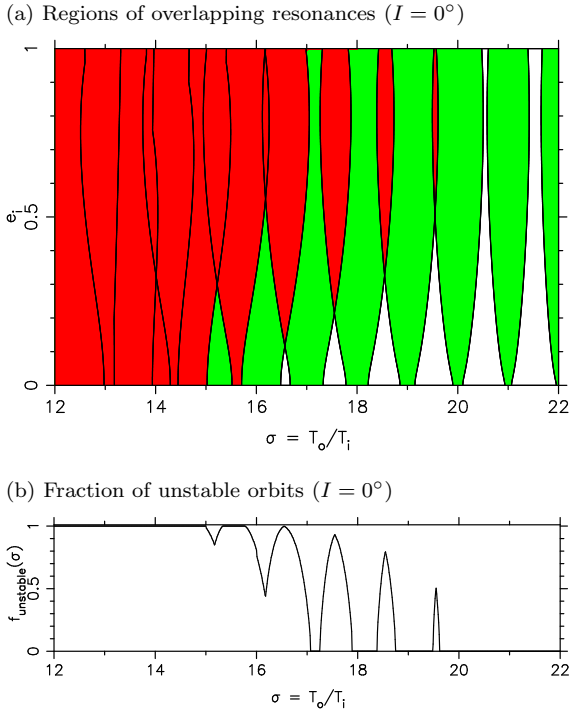

\begin{centering}$\begin{array}{c}
\multicolumn{1}{l}{\mbox{(a) Regions of overlapping resonances ($I=0\deg$)}}\\[-0.1cm]
\includegraphics[height=\figsmall,angle=270]{ResPlot_RG.ps}\\
\multicolumn{1}{c}{\mbox{}}\\
\multicolumn{1}{l}{\mbox{(b) Fraction of unstable orbits ($I=0\deg$)}}\\[-0.1cm]
\includegraphics[height=\figsmall,angle=270]{FUnstable_e5I0.ps}
\end{array}$
\par\end{centering}
\caption[Stability of coplanar orbits in a globular cluster]
{Resonance widths for a particle orbiting the cluster core with a perturbing galaxy on a coplanar $e_o=0.5$ orbit for $m_2/m_1 = 10^{-6}$ and $m_3/m_1 = 10^5$. Regions of resonance overlap are shaded in red to show the predicted unstable orbits in panel (a), while regions inside a single resonance width are shaded in green. Panel (b) shows the predicted fraction of unstable orbits as a function of the ratio of outer to inner orbital periods after averaging over the distribution of eccentricities in a cluster (shown in Figure~\ref{gc_distEcc}).}
\label{gc_SB_Shaded}
\end{figure}

To demonstrate the stability analysis process the MSC is applied to a co-planar system with outer eccentricity $e_o = 0.5$ following the method given in Section~\ref{MSC}. The resonance widths for $n:1$ type resonances are determined from \reqOne{th_ResWidth} and are shown in Figure~\ref{gc_SB_Shaded} (a) as a function of $\sigma = T_o/T_i$ for $m_2/m_1 = 10^{-6}$ and $m_3/m_1 = 10^5$. 

Regions where the system resides in a single resonance are shaded green in Figure~\ref{gc_SB_Shaded} (a) and the boundary of this region (the separatrix) is indicated by a black curve. The resonance width calculated by \reqOne{th_ResWidth} is the distance of the separatrix from exact resonance ($n:1$). Regions where two or more resonances overlap are shaded in red to indicate theoretically unstable orbits. In the context of globular clusters, stars on these orbits are expected to eventually escape from the cluster.

Unstable systems can still occur near the separatrix \citep[see Figure 15 of][]{RoIoA}, which means that the predicted unstable regions are a conservative estimate. The criterion of an unstable system being any system that resides in two resonances simultaneously is adopted as a quick diagnostic and gives a good estimate as to where most unstable regions of $\sigma$-$e_i$ space occur. 

By summing along the inner eccentricity weighted by the eccentricity distribution function for stellar particles in a Plummer sphere given by \reqOne{gc_CummEcc} the fraction of unstable orbits as a function of $\sigma$ can be determined. By introducing a stability function $S(\sigma,e_i)$ which is 1 if the system is unstable and 0 otherwise, then the fraction of unstable orbits can be written as
\begin{equation}
f_{unstable}(\sigma) = \int_0^{1} f(e_i) S(\sigma, e_i) de_i
\label{gc_funs_coplanar}
\end{equation}
where $f(e_i)$ is given by \reqOne{gc_eccdistfn}. The fraction of unstable orbits for the co-planar $e_o = 0.5$ system is shown in Figure~\ref{gc_SB_Shaded} (b) using the stability values from panel (a).

\begin{figure}
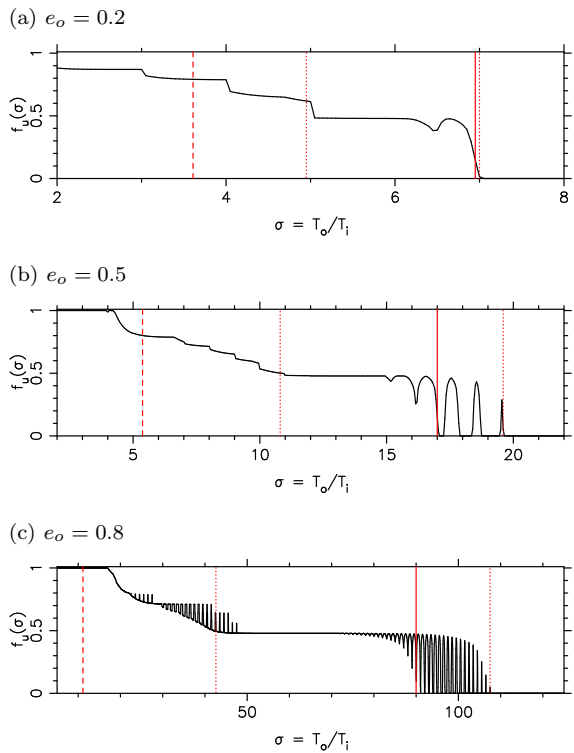

\begin{centering}$\begin{array}{cc}
\multicolumn{1}{l}{\mbox{(a) $e_o = 0.2$}}\\[-0.1cm]
\includegraphics[height=\figsmall,angle=270]{FUnstable_e2.ps}\\
\multicolumn{1}{c}{\mbox{}}\\
\multicolumn{1}{l}{\mbox{(b) $e_o = 0.5$}}\\[-0.1cm]
\includegraphics[height=\figsmall,angle=270]{FUnstable_e5.ps}\\
\multicolumn{1}{c}{\mbox{}}\\
\multicolumn{1}{l}{\mbox{(c) $e_o = 0.8$}}\\[-0.1cm]
\includegraphics[height=\figsmall,angle=270]{FUnstable_e8.ps}
\end{array}$
\par\end{centering}
\caption[Effect of outer eccentricity on unstable fraction]
{The effect of the outer eccentricity on the fraction of unstable orbits ($f_u(\sigma)$) against the period ratio ($\sigma$). The unstable fraction is averaged over the inner eccentricity and the relative inclination by Equation~\ref{gc_funs_inc}. The transition from unstable to stable orbits $\sigma_u$ is shown as a vertical dashed line and is the lowest $\sigma$ value where the fraction of unstable orbits drops beneath 10\%. The minimum and maximum $\sigma$ values associated with the width of this transition are shown as dotted lines while the King radius is shown as a dashed line. Note the increase in the range of $\sigma$ as $e_o$ increases.}
\label{gc_Funstable_eout}
\end{figure}

The final fraction of unstable orbits as a function of the period ratio $\sigma$ is determined by averaging \reqOne{gc_funs_coplanar} across the range of relative inclinations between the star-cluster and cluster-galaxy orbits. For orbits with relative inclination $I$ the inclination terms listed in Section~\ref{MSC} are used to calculate the resonance width and hence the stability of orbits. The final fraction of unstable orbits can be written as
\begin{equation}
f_{u}(\sigma) = \int_0^{2\pi} g(I) f_{unstable}(\sigma, I) dI
\label{gc_funs_inc}
\end{equation}
where for a uniform sphere the distribution function for inclination is given by
\begin{equation}
g(I) = \frac{1}{2 \pi} \left( 1 + \cos 2 I \right).
\label{gc_ProbDistFn_Inc}
\end{equation}
In practice both equations~(\ref{gc_funs_coplanar}) and~(\ref{gc_funs_inc}) are calculated numerically using a resolution of $\Delta e_i = 0.005$, $\Delta_I = 10\deg$ and $\Delta \sigma = 0.05$. The fraction of unstable orbits after averaging over the relative inclination and eccentricities of stellar orbits within a cluster, as determined by \reqOne{gc_funs_inc}, is shown in Figure~\ref{gc_Funstable_eout} (a), (b) and (c) for $e_o = 0.2$, 0.5 and 0.8 respectively. 

To characterise the stability boundary three period ratio values are used, all of which are found numerically when calculating $f_u(\sigma)$. Firstly there is the representative value $\sigma_u$ which is defined as the lowest $\sigma$ value where the fraction of unstable orbits drops beneath 10\%, this is shown as the red solid line in Figure~\ref{gc_Funstable_eout}. The maximum period ratio $\sigma_{max}$ is defined as the highest $\sigma$ value for which $f_{u}(\sigma) > 10^{-3}$ and represents the deepest into the cluster that stars can be on unstable orbits. The final value $\sigma_{min}$ is defined as the lowest period ratio where $f_{u}(\sigma) < 0.5$, this criterion is used to divide the unstable region into two categories. The first category is represented by $\sigma < \sigma_{min}$ which is where all stars are unstable to escape from the cluster, and the second by $\sigma_{min} < \sigma < \sigma_u$ where stars on prograde orbits are preferentially removed. It is well established that stars on retrograde orbits are more stable to escape than prograde orbits \citep[e.g.][]{RWEGK2006}, which is expected from the inclination terms in the MSC given in \reqTo{th_incStart}{th_incEnd}. The minimum and maximum period ratios are shown as dotted lines in Figure~\ref{gc_Funstable_eout} and the period ratio associated with the King radius (Equation~\ref{rtidal_king1}) is shown as dashed lines. These three period ratios characterise the transition from stable orbits inside the cluster to unstable orbits in the outer regions of the cluster, as illustrated in Figure~\ref{gc_3body_concept}.

From Figure~\ref{gc_Funstable_eout} the transition from unstable to stable orbits ($\sigma_u$) increases as outer eccentricity ($e_o$) increases. This is due to the dependence of the resonance width on the combination $n \xi(e_o)$ in \reqOne{th_ResFn_eo}, where $n$ refers to the $n:1$ resonance. This quantity is always positive and as it increases the resonance width rapidly falls to zero. Unstable systems therefore require $n \xi(e_o)$ to be as close to zero as possible, which is achieved for high values of $n$ (and $\sigma$) if $e_o$ is also high. Physically, this reflects the fact that an exponentially small amount of energy is exchanged between the inner (star-cluster) and outer (cluster-galaxy) orbits when their orbits are very wide \citep{RoIoA}.

The width of the transition from unstable to stable orbits also increases with $e_o$ as seen in the progression of panels (a) through (c) of Figure~\ref{gc_Funstable_eout}. Note that the basic structure of peak stability occurring at integer values of $\sigma$ and stability increasing with increasing $\sigma$ is consistent for all eccentricities. This phenomenon is expected since resonance widths from the $n:1$ and $n+1:1$ resonances (where $n<\sigma<n+1$) overlap at the midpoint between these resonances, as seen previously in Figure~\ref{gc_SB_Shaded} (a) for the coplanar case.

\begin{figure}
\begin{centering}\resizebox{\figsmall}{!}{\includegraphics[angle=270]{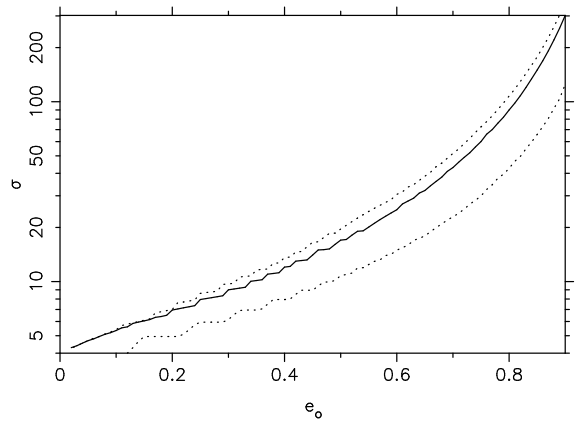}} 
\par\end{centering}
\caption{The effect of the galactic orbital eccentricity of the cluster on the transition from unstable to stable star-cluster orbits. The transition $\sigma_u$ is shown as a solid line and is used as an estimate for the stability radius, while the associated transition width values $\sigma_{min/max}$ are shown as dotted lines.}
\label{fig:sige}
\end{figure}

The results for all eccentricity values are shown in Figure~\ref{fig:sige}, from which a near exponential dependence of $\sigma_u$ on $e_o$ is seen. This is also true for the width of the transition between unstable and stable orbits as represented by $\sigma_{min/max}$ and shown as dotted curves on each side of $\sigma_u$ in Figure~\ref{fig:sige}. Note that the coarseness in the curves for all $\sigma$ values for low eccentricities is due to a resolution of $0.1 \sigma$ used when producing this figure. For high eccentricities it is clear from the figure that there is a large range in period ratios from which stars can potentially escape the cluster.

The transition value of $\sigma_{u}$ is used in Section~\ref{Comparison} to calculate the stability radius for a given perigalacticon and eccentricity of the cluster orbit about the galaxy. The width of the transition from unstable to stable orbits, given by $\sigma_{min/max}$, is used to provide approximate error bars associated with the stability radius. The approximate dependence of the stability boundary period ratio on the eccentricity is found by fitting all data points in Figure~\ref{fig:sige}, giving
\begin{equation}
\sigma(e_o)^{-3/7} = 0.03869 \exp \left( -6.3218 e_o \right)
\label{sigeccdep}
\end{equation}
where the form is chosen for convenience when calculating the relaxation timescale in Section~\ref{gc:valid}.

This stability radius calculation is intended to be generally applicable to the entire system of globular clusters. However in determining $\sigma(e_o)$ the mass ratios have been taken as constant. This is implicitly reflected in the stellar cluster model used to determine the probability distribution for the eccentricity of stellar orbits within the cluster (\reqNP{gc_CummEcc}). This cluster consisted of $1 \msun$ particles in a cluster of mass $M_C = 10^6 \msun$, which will not be true of all clusters. 

Changing the cluster mass is expected to have a small effect on the resonance width calculation and therefore on the stability boundary as a function of eccentricity. This can be demonstrated by considering the mass dependence of \reqOne{th_AInc} when $m_2 = m_i \ll M_C = m_1$ and $m_1 = M_C \ll M_G = m_3$, which can be simplified to
\begin{equation}
\Delta \sigma \propto \sqrt{ 1 + \left( \frac{a_o}{a_i} \right) \left( \frac{M_C}{M_G} \right)^{2/3} } 
\end{equation}
where for distant globular clusters $a_o \sim 10$ kpc, $a_i \sim 10$ pc and $M_G = 10^{11} \msun$ means that this term is effectively independent of mass for $10^{4} \lapprox M_C/\msun \lapprox 10^6$. Since the resonance width does not depend on the choice of $m_i$ the numerical and analytical results presented here are applicable to real GCs with a mean stellar mass of $m_i \approx 0.4 \msun$. We can therefore apply the $\sigma(e_o)$ relationship derived in this section to all cluster masses for distant globular clusters.

\section{Expected region of validity}
\label{gc:valid}

A region in GC-galaxy orbital phase space is sought such that three-body instability occurs on a timescale shorter than the age of the GC and the relaxation timescale. To this end the time taken for a star to escape the cluster is determined from the numerical integrations presented in Section~\ref{stability_test}. These are then compared to the calculated relaxation timescales at the stability boundary to see if sufficient time is available for the star to escape.

\begin{figure}
\begin{centering}\resizebox{\figsmall}{!}{\includegraphics[angle=270]{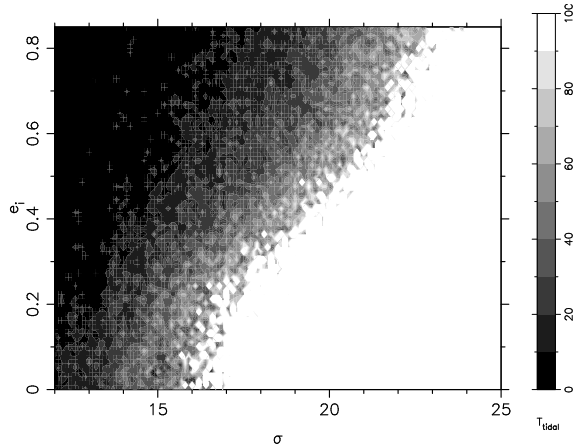}} 
\par\end{centering}
\caption{Timescale for the stellar particle to escape beyond two times the tidal (King) radius as a function of period ratio and inner eccentricity. Each point represents the average of 10 random realisations of the initial phase of the orbit. Timescales for the case with relative inclination of $30\deg$ are shown here in units of the outer period ($T_o$); similar results were found for $I=0\deg$ and $I=60\deg$.}
\label{EscapeTimes}
\end{figure}

\begin{figure}
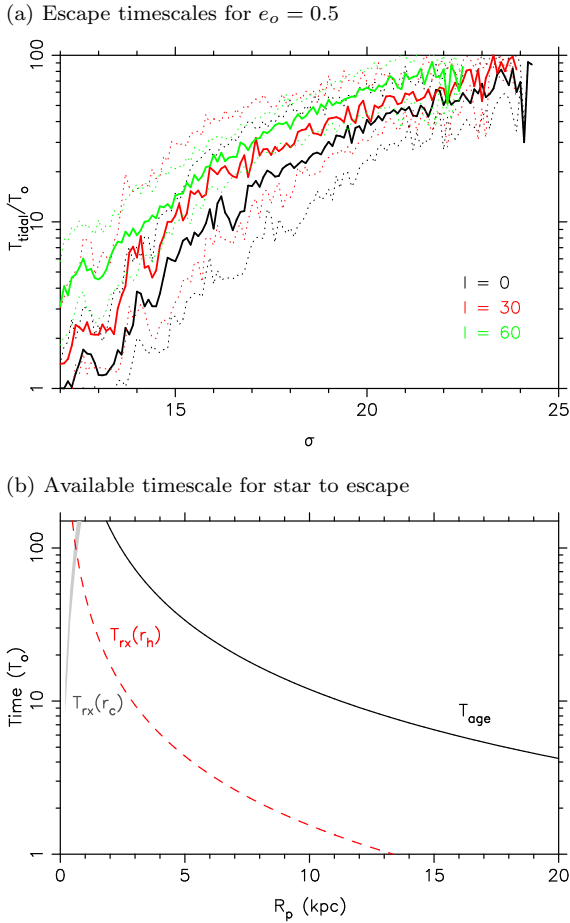

\begin{centering}$\begin{array}{c}
\multicolumn{1}{l}{\mbox{(a) Escape timescales for $e_o = 0.5$}}\\[-0.1cm]
\includegraphics[height=\figsmall,angle=270]{TimeTidal.ps}\\
\multicolumn{1}{c}{\mbox{}}\\
\multicolumn{1}{l}{\mbox{(b) Available timescale for star to escape}}\\[-0.1cm]
\includegraphics[height=\figsmall,angle=270]{RperiTidal.ps}
\end{array}$
\par\end{centering}
\caption
{Top panel shows the average timescale for a star to exceed the King tidal radius ($2 R_K$) against the period ratio $\sigma$ for stars that are unstable to escape. The timescales ($T_{tidal}$) are averaged over the star-cluster orbital eccentricities using the numerical results for $e_o=0.5$ taking $r_h = 4$ pc and $M_C = 10^6 \msun$. The inclination dependence is shown with $I = 0\deg$, $30\deg$ and $60\deg$ as black, red and green curves respectively and the associated 68\% confidence regions are indicated by dotted lines. The bottom panel shows the half-mass relaxation timescale (red dashed line), relaxation timescale for the stability boundary defined as $\sigma_{min}<\sigma<\sigma_{max}$ (grey region) and the approximate age of the galactic GC system (black curve) against the perigalacticon, assuming $e_o = 0.5$ and a Keplerian galactic potential.}
\label{fig_sigperi}
\end{figure}

The balance between the escape timescale and the relaxation timescale is important because the three-body instability is a resonance effect so it will be disrupted by the random kicks from two-body encounters. As the MSC can only predict the occurrence of unstable orbits and cannot be used to determine the associated timescale numerical experiments must be used. Firstly, $T_{tidal}$ is defined to be the time taken for a star to escape beyond $2 R_{K}$ from the cluster particle; given that the star did escape before the end of the simulation. Secondly, the previous numerical experiments (Section~\ref{stability_test}) are used to show this timescale as a function of $\sigma$ and $e_i$ averaged over 10 randomly selected phase angles. The resulting timescales in units of the outer period are shown in Figure~\ref{EscapeTimes} for a relative inclination of $30\deg$; note that similar results were found for $I=0\deg$ and $I=60\deg$. The key result is that for unstable regions of $\sigma$ and $e_i$ the timescale is typically less than 10 outer orbits, but can be as high as $100 T_o$ near the stability boundary. 
Using the eccentricity averaging method described in Section~\ref{EccDist} the average escape timescale can be determined from the numerical results for each inclination. The median escape timescales against the period ratio for each inclination are shown in Figure~\ref{fig_sigperi} (a). The associated 68\% confidence interval around the median values are indicated by dotted curves for each inclination.

Two-body relaxation is an internal process and as such does not depend on the GC-galaxy orbit, only on the internal structure of the cluster. This means that a half-mass radius must be chosen to compare the relaxation timescale to the escape timescale. The two-body relaxation time at the half-mass radius is given by \cite{Spitzer1987} 
\begin{equation}
T_{rx}(r_h) = \frac{0.14 N}{2 \pi \ln(0.4 N)} T_i(r_h)
\label{gc_relaxtime}
\end{equation}
where $T_i(r_h)$ is the orbital period of star with semi-major axis given by the half-mass radius and we choose $r_h = 4$ pc. The relaxation timescale at the half-mass radius in units of the period of the GC-galaxy orbit ($T_o$) against perigalacticon ($R_p$) is shown as the red dashed line in Figure~\ref{fig_sigperi} (b). For a cluster mass of $10^6 \msun$ the relaxation timescale is approximately 1.3 Gyr at the half-mass radius, but it is the timescale at the stability boundary that is of interest there, so the radial dependence for the relaxation timescale is required.

Describing the GC as a Plummer sphere then the density profile in the outer regions is $\rho \propto r^{-5}$ and the velocity dispersion is $v \propto r^{-1/2}$ so the radial dependence of the relaxation timescale is approximated by
\begin{equation}
T_{rx}(r) \propto \frac{v^3}{\rho} \propto r^{7/2}.
\label{trxraddep}
\end{equation}
Assuming a star near the stability boundary is sufficiently distant from the cluster centre such that the period is Keplerian, then $T_i \propto r^{3/2}$ and $\sigma = T_o/T_i \propto r^{-3/2}$ so Equation~\ref{trxraddep} can be written in terms of the period ratio at the half-mass radius ($\sigma_h$) as
\begin{equation}
T_{rx}(\sigma) = T_{rx}(r_h) \left( \frac{\sigma}{\sigma_h} \right)^{-7/3}
\end{equation}
whereby assuming the outer period is also Keplerian then
\begin{equation}
\frac{T_{rx}(\sigma)}{T_{rx}(r_h)} = \left( \frac{M_C}{M_G} \right)^{7/6} \left( \frac{R_p}{r_h} \right)^{7/2} \left( 1 - e_o \right)^{-7/2} \sigma^{-7/3}.
\label{gc_relaxtimesig}
\end{equation}
The period ratio at the stability boundary is constant for a given eccentricity (see Figure~\ref{fig:sige}) and so the relaxation timescale at the stability boundary increases rapidly with perigalacticon. The stability region ($\sigma_{min} < \sigma < \sigma_{max}$) is shown as the shaded grey region in Figure~\ref{fig_sigperi} (b). The timescale for a star to random walk out of the cluster due to three-body instability is less than the two-body relaxation time in the regions below grey region. The final constraint shown in Figure~\ref{fig_sigperi} (b) is the approximate age of the Milky Way globular cluster system, taken as $T_{age}=10$ Gyr, which imposes an upper limit on how many GC-galaxy orbits are possible.

By comparing the amount of time needed for a star to escape the cluster (Figure~\ref{fig_sigperi} a) with the amount of time available (Figure~\ref{fig_sigperi} b) the regime where chaotic effects are important can be determined. There is a wide range of perigalacticon values where stars are subject to escape by three-body instability; namely $0.5 \lapprox R_p/$kpc$\lapprox 11$ for an escape time of $T_{tidal} = 10 T_o$. The lower limit of this range is set by the relaxation timescale at the stability boundary and the upper limit is set by the age of the galactic GC system.

The results presented in Figure~\ref{fig_sigperi} are specific to an GC-galaxy orbital eccentricity of 0.5 so the effect of changing the eccentricity on the perigalacticon range must be examined. At the upper limit the same constraint based on the available time for the GC-galaxy orbit still applies. While at the lower limit the effect of eccentricity on the relaxation timescale and the stability boundary is determined by substituting the stability boundary dependence on eccentricity given by Equation~\ref{sigeccdep} into Equation~\ref{gc_relaxtimesig} giving
\begin{equation}
T_{rx}(e_o) \propto \frac{0.03869 \exp \left( -6.3218 e_o \right) }{\left( 1 - e_o \right)^{7/2}}
\label{trxsige}
\end{equation}
which has a local minimum at $e_o = 0.45$, close to the chosen eccentricity value of 0.5, so Figure~\ref{fig_sigperi} (b) shows the worst case scenario. In other words all other eccentricities will increase the relaxation timescale at the stability boundary and therefore increase the perigalacticon region where three-body stability is significant.

In summary two key results have been found, firstly that the timescale for a star to escape the cluster was found to be of the order of 10 GC-galaxy orbital periods. Secondly that for these half-mass radius, $M_C$ and $e_o$ values, the timescale associated with three-body instability leading to stars crossing the tidal radius is sufficiently rapid for GC-galaxy orbits with $R_p$ greater than 1 kpc. As there is sufficient time for stars to escape the cluster via the three-body instability for most, if not all, GCs in the Milky Way system we can proceed to apply the stability boundary to real GC orbits.

\section{Comparison with previous theoretical work}

\label{Comparison}

Two theoretical estimates for the tidal radius from the literature are compared to the stability radius derived from the MSC method. The first estimate from the literature is the most commonly used tidal radius estimate in the field derived by \citet{KingI} and will be referred to as the King radius. The second is an extended analytical determination that was also compared to N-body simulations for a single set of orbital parameters, this determination is given in \citet{RWEGK2006} and will be referred to as the Read radius.

To aid comparison between these estimates the eccentricity dependence is separated out so that the tidal radius can be written as 
\begin{equation}
r_t = R_p \left( \frac{M_C}{M_G} \right)^{1/3} f(e_o)
\label{rtidal}
\end{equation}
where $M_C$ and $M_G$ are the masses of the cluster and galaxy respectively, $e_o$ is the eccentricity of the clusters orbit around the galaxy, and $R_p$ is the distance of closest approach to the galaxy, referred to as the perigalacticon.

The simplest case to determine analytically is to consider a star located where the acceleration on the star in the rotating frame is zero and the velocity of the star relative to the cluster centre is also zero. Such a star will be on a radial orbit with respect to the centre of the cluster. For a star on a radial orbit and using point mass potentials for the cluster and galaxy the eccentricity dependence of the tidal radius is given by \citep{KingI}
\begin{equation}
f(e_o) = k \left( 3 + e_o \right)^{-1/3}
\label{rtidal_King}
\end{equation}
where if $k=1$ the Jacobi radius is recovered. Here the constant $k = 0.7$ is used, as was introduced by \citet{Keenan1981} to better fit observations of the galactic globular clusters. 

\citet{RWEGK2006} extended the tidal radius calculation of \citet{KingI} by including the Coriolis terms in the equations of motion, which takes the effect of the orbit of the star into account. To simplify the equations of motion they limited their analysis to co-planar systems with stars on prograde or retrograde circular orbits or purely radial orbits. They found that a star has zero acceleration in the rotating cluster frame if the eccentricity dependence of the distance from the cluster is given by \citep{RWEGK2006} 
\begin{eqnarray}
f(e_o)= \left( \frac{1}{1+e_o} \right)^{1/3} \left( \frac{\sqrt{\alpha^2 + 1 + \frac{2}{1+e_o}} - \alpha}{1+\frac{2}{1+e_o}} \right)^{2/3}
\label{rtidal_Read}
\end{eqnarray}
where $\alpha = 0$ denotes a star on a purely radial orbit with the cluster centre in the rotating cluster frame and reduces to \reqOne{rtidal_King} in this case. The analysis by \citet{RWEGK2006} also considered cases where the radial velocity is zero at the point of zero acceleration, which occurs stars on circular orbits and is described in the tidal radius equation by setting $\alpha = 1$ or $-1$ for prograde and retrograde orbits respectively (note that $\alpha = -1$ gives $r_t$ greater than the King radius). For later comparison we define the Read radius as \reqOne{rtidal_Read} with $\alpha = 1$, representing an easily tidally stripped cluster. The Read radius compared well with two N-body simulations of $10^7 \msun$ satellite clusters using $N = 10^5$ particles with orbital parameters of perigalacticon $R_p/R_{1/2} = 267$ and eccentricity $e_o = 0.0$ and perigalacticon $R_p/R_{1/2} = 77$ and eccentricity $e_o = 0.57$ \citep{RWEGK2006}. A summary of alternate equations for the tidal radius designed to fit observations can be found in \citet{Bellazzini2004}. 

\begin{figure}
\begin{centering}\resizebox{\figsmall}{!}{\includegraphics[angle=270]{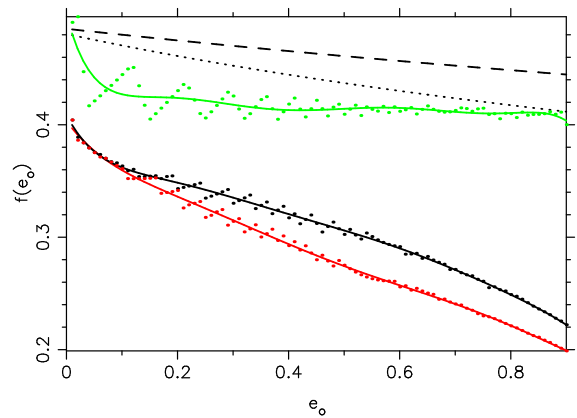}} 
\par\end{centering}
\caption
{The eccentricity dependence of the stability radii associated with the MSC prediction are shown as data points along with a solid line fit given by \reqOne{rtidal_MSC}, the King radius as dashed lines and the prograde Read radius as dotted lines. The retrograde Read radius ($\alpha = -1$) is not shown since $f(e_o) \sim 1$ for all eccentricites. Coefficients to fitting functions are given in Table~\ref{Coefficients} along with the minimum (red) and maximum (green) fits.}
\label{fig:rtidal}
\end{figure}

For the stability boundary determined in Section~\ref{MSC} to be useful it must first be converted into an equivalent radius from the cluster centre. Assuming that the gravitational potential of the cluster is well approximated by a point mass at distances of the tidal radius for clusters of interest then the stability radius can be written in the same form of \reqOne{rtidal}. By converting the period ratio shown in Figure~\ref{fig:sige} into an equivalent semi-major axis ($a_i$) via $T_o = \sigma T_i$ and using the resulting $a_i$ as $r_t$ then the eccentricity dependence itself as defined by $f(e_o)$ in \reqOne{rtidal} can be determined. The values for $f(e_o)$ as determined from the period ratio $\sigma$ dependence on eccentricity (Figure~\ref{fig:sige}) are shown as the data points in Figure~\ref{fig:rtidal}. These data can be fit by
\begin{equation}
f(e_o) = \exp \left[ \sum_{i=0}^{N=7} a_i (e_o)^i \right]
\label{rtidal_MSC}
\end{equation}
where the coefficients are given in Table~\ref{Coefficients} for the minimum, maximum and indicative (unstable) radii. Note that the maximum boundary is also well approximated simply by $f_{max} \approx 0.42$, especially for higher eccentricities.

\begin{table}
\begin{centering}\begin{tabular}{|l|l|l|l|}
\hline 
 & $f_{u}$ & $f_{min}$ & $f_{max}$ \\
\hline 
$a_0$	&	-0.89462	&	-0.907914	&	-0.696568	\\
$a_1$	&	-2.36353	&	-1.67172	&	-3.83438	\\
$a_2$	&	17.6103	&	7.2727	&	36.6515	\\
$a_3$	&	-77.3899	&	-23.8429	&	-177.338	\\
$a_4$	&	185.385	&	36.8088	&	461.984	\\
$a_5$	&	-247.308	&	-22.2994	&	-657.971	\\
$a_6$	&	171.836	&	-1.80097	&	482.319	\\
$a_7$	&	-48.6078	&	4.73382	&	-142.327	\\
\hline
\end{tabular}\par\end{centering}
\caption{Coefficients for fit to $f_{u}(e_o)$ and the minimum and maximum extents of the marginally chaotic zone. For radii with $f(e_o)< f_{min}(e_o)$ all orbits are expected to be stable, whereas for $f(e_o) > f_{max}(e_o)$ they are expected to all be chaotic.}
\label{Coefficients} 
\end{table}

Recall from Section~\ref{GCRTidal} that stars on prograde orbits are predicted to be preferentially removed relative to stars on retrograde orbits. This was also seen in Figure~\ref{gc_Funstable_eout} where only prograde orbits are unstable in the region between $\sigma_{min}$ and $\sigma_{u}$. Therefore more stars with retrograde orbits exist in the outer regions on the cluster, leading to a net rotation being predicted between the indicative radii ($f_u$) and the maximum radii ($f_{max}$).

From Figure~\ref{fig:rtidal} we see that the boundary between stable and unstable orbits occurs interior to both the Read and King tidal radius estimates. However the stability boundary is not equivalent to the tidal radius. This is because although chaotic orbits will eventually result in the escape of the star from the globular cluster, this occurs via a random walk process. Therefore there will be stars remaining outside the stability boundary for approximately 10 GC-galaxy orbits; under various assumptions of the GC orbit and cluster half-mass radius (see Section~\ref{gc:valid}). 

By increasing the binding energy of stars in the outer regions of a cluster, chaotic diffusion will contribute to the `potential escapers' population of stars discussed by \citet{FH2000}. They found that there exists a population of stars with energy greater than the potential energy at the tidal radius whose escape timescale can be longer than the cluster age. \citet{KupperEtAl2010} found using N-body simulations that the potential escapers cause a smoothing out of the velocity dispersion inside of the tidal radius. This difference in the velocity dispersion to what is expected for a relaxed and isolated globular cluster forms the basis for the second paper in this series \citep{PaperII}.

For some globular clusters, stars that should theoretically be removed by the random walk in binding energy associated with three-body instability will not escape due to this process being suppressed by two-body relaxation. The perigalacticon range for the GC-galaxy orbits where this is expected to occur has been shown in Figure~\ref{fig_sigperi} for $e_o =0.5$ with a half-mass radius of 4 pc. The ratio of the escape timescale to the relaxation timescale is further examined in \cite{PaperII} when the stability boundary is determined for a sub-sample of Milky Way globular clusters. In addition, Kennedy (in preparation) will investigate the escape process in far more detail using a high resolution N-body simulation in a realistic galactic potential for many galactic orbits.

\section{Discussion and conclusions}
\label{Conc}

The stability radius determined here differs from previous tidal radius estimates in that it emerges naturally from a stability analysis of the general three-body problem. It was found that the eccentricity of the cluster-galaxy orbit is expected to have a far more significant effect on the tidal radius of globular clusters than previous theoretical results have found. 

The approach adopted here means that what is actually predicted is the boundary between stable and unstable orbits for stars inside a cluster, which is interior to previous tidal radius values for all parameters of the cluster-galaxy orbit. This was found to be the case in comparison to the commonly used tidal radius by \citet{KingI} and to the more recent radius by \citet{RWEGK2006}. One key difference between the stability radius and previous tidal radius estimates is that a much stronger dependence on the eccentricity of the cluster-galaxy orbit is predicted here. 

This eccentricity dependence has been studied using N-body simulations by \citet{KupperEtAl2010} and \citet{WebbEtAl2013} who found that the limiting radius is larger than the classical King radius. However these studies did not examine the chaotic diffusion process and the timescale for how long stars on unstable orbits take to escape the cluster. A simulation focussed on the chaotic diffusion process is ongoing and will appear in a future publication.

From a practical point of view the major outcome of this work is the derivation of an easy to use stability radius of the form
\begin{equation}
r_t = R_p \left( \frac{M_C}{M_G} \right)^{1/3} f(e_o)
\end{equation}
where $f(e_o)$ is given by a function (\reqNP{rtidal_MSC}) fitted to stability results determined analytically from the stability of the general three-body problem. This has a wide range of applications, including the effect of unstable orbits on the velocity dispersion profile for Milky Way globular clusters, which is taken up in the second paper of this series \citep{PaperII}.

\section{Acknowledgements}

I am indebted to Rosemary Mardling for the proposing the idea behind this paper as part of my thesis work and for supplying the unpublished inclination terms used in this study. I would also like to thank Alan Bond and the Monash Electrochemistry group who supported me on another project while I was carrying out this investigation. Computations of three-body orbits were carried out using NIMRODG which was developed and maintained by the Monash eScience and Grid Engineering Laboratory ({MeSsAGE} Lab). I also acknowledge the support by the Chinese Academy of Sciences for young international scientists, grant number O929011001.

\bibliography{GCR.bib}

\end{document}